\documentclass{aastex}          
\usepackage{spr-astr-addons}    

\begin{document}
%
\title{The young star cluster system of the Antennae galaxies}

\shorttitle{YSCs in the Antennae galaxies}
\shortauthors{Peter Anders}

\author{Peter Anders\altaffilmark{1}} 
\and 
\author{Uta Fritze\altaffilmark{2}}
\and 
\author{Richard de Grijs\altaffilmark{3}}

\altaffiltext{1}{Sterrenkundig Instituut, Universiteit Utrecht, P.O. Box 80000, NL-3508 TA Utrecht, The Netherlands}
\altaffiltext{2}{Centre for Astrophysics Research, University of Hertfordshire,
College Lane, Hatfield AL10 9AB, UK}
\altaffiltext{3}{Department of Physics \& Astronomy, The University of Sheffield,
Hicks Building, Hounsfield Road, Sheffield S3 7RH, UK}

\begin{abstract}

The study of young star cluster (YSC) systems, preferentially in
starburst and merging galaxies, has seen great interest in the recent
past, as it provides important input to models of star formation.
However, even some basic properties (like the luminosity function
[LF]) of YSC systems are still under debate. Here we study the
photometric properties of the YSC system in the nearest major
merger system, the Antennae galaxies. We find evidence for the
existence of a statistically significant turnover in the LF.

\end{abstract}


\section{Introduction}
\label{s:intro}

Star clusters (SCs) form nearly instantaneously through the collapse
of giant molecular gas clouds. Hence, all stars within a SC are
approximately coeval, share the same chemical composition, and
therefore represent a simple stellar population. A small number of
parameters, in particular their initial chemical composition and
initial stellar mass function, are enough to describe their colour and
luminosity evolution on the basis of a given set of stellar
evolutionary tracks or isochrones (e.g.
\citealt{1999ApJS..123....3L,2003A&A...401.1063A,2003MNRAS.344.1000B}).
Therefore, observed spectrophotometric properties of SCs are
relatively easy and straightforward to interpret. 

SC formation is a major mode of all star formation, and possibly even
the dominant mode in strong starbursts triggered in gas-rich galaxy
mergers (e.g., \citealt{1995Natur.375..742M,2003NewA....8..155D}). In
addition, as SCs inherit and conserve the chemical abundances at the
place and time of their birth up to old ages, they are excellent
tracers of their parent galaxy's properties in terms of star formation
and chemical enrichment history. 

One of the most basic and commonly used diagnostics to explore the
properties of entire SC {\sl systems} is their LF. While for old
globular cluster systems the Gaussian shape of their LFs (and indeed
mass functions) is well established (see e.g.
\citealt{1998gcs..book.....A,1991ARA&A..29..543H}), the situation for
young SCs is still under discussion. While mainly LFs consistent with
a power-law, resembling the MF of nearby molecular cloud cores, are
quoted for YSC systems ranging from Galactic open cluster to YSCs in
major mergers (e.g.
\citealt{1984AJ.....89.1822V,2003AJ....126.1836H,1998AJ....116.2206S,1999AJ....118.1551W},
see \citealt{2003MNRAS.343.1285D} for a recent compilation), some
studies find deviations from a power-law or direct evidence for
Gaussian distributions
(\citealt{2006MNRAS.366..295D,1999A&A...342L..25F,2004ApJ...613L.121G}).

To contribute to our understanding of this astrophysical question, which ties
directly to the fundamental physical conditions of star formation in general, we
performed an analysis of the young ($\sim 0 - 100$ Myr) SC system formed during
the, still ongoing, merging process in the nearest major merger of two giant
gas-rich spiral galaxies, the Antennae galaxies, i.e. NGC 4038/39.

\section{Photometry} 
\label{s:photo} 

We reanalysed the most homogeneous broad-band dataset including the U
band of the Antennae galaxies available (providing photometry in
$UBVI$), obtained using {\sl HST}/WFPC2 as part of programme GO-5962
(PI B. Whitmore).

We select our clusters based on:
\begin{itemize}
\item a flux threshold compared to its surrounding
\item cross-correlation between filters (i.e. a cluster must be detected in several
bands)
\item photometric errors $<$ 0.2 mag in each band
\item size measurements using {\sc BAOlab} (\citealt{1999A&AS..139..393L})
\end{itemize}

Selecting only clusters in the (PSF-corrected) size range $R_{1/2} \simeq 5$ to 25
pc (to avoid contamination by single stars and cluster complexes, respectively)
yields a sample of 365 clusters, satisfying all selection criteria. Utilising the
cluster sizes we apply size-dependent aperture corrections, as described in
\cite{2006A&A...451..375A}.

\section{Completeness} 
\label{s:complete} 

In this study we are primarily interested in the LF of the young star
clusters. A consistent and reliable determination of the completeness
fraction as function of cluster magnitude {\bf and all} other relevant
selection criteria/effects is therefore of prime importance.

\begin{figure}[t]
\includegraphics[angle=270,width=\columnwidth]{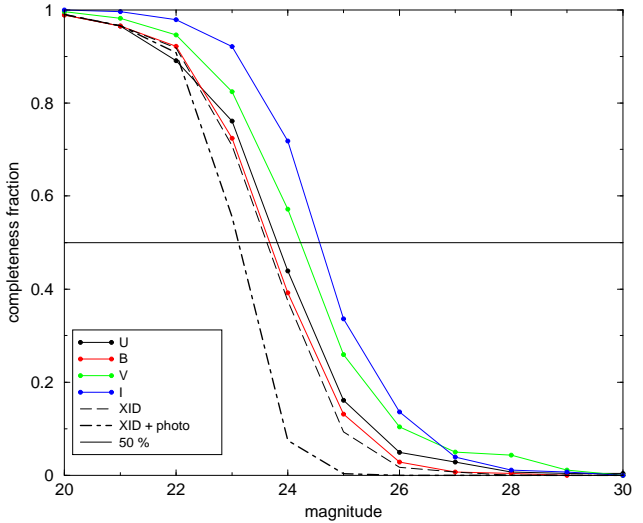}
\caption{Completeness fractions for artificial clusters with FWHM = $\simeq$ 10 pc
for a variety of selection criteria (see text for details)} 
\label{fig:complete}
\end{figure}

We investigated the completeness for a number of restrictions, taking
successively more cluster selection criteria into account. Some of
these results for artificial clusters with FWHM = $\simeq$ 10 pc in a
cluster-rich region in NGC 4038 are shown in Fig. \ref{fig:complete}.

First, we determined the completeness in each band independently
(labelled "U","B", "V" and "I" in Fig. \ref{fig:complete}). Taking the
cross-correlation into account (labelled "XID") reduced the
completeness to slightly below the most limiting single-band
completeness. By applying the photometric uncertainties restriction
(labelled "XID + photo") the 50\% completeness magnitude is decreased
by about $\simeq 0.4$ mag.

Using {\sl stars} instead of extended objects as test sources
artificially increases the 50\% completeness magnitude by $\simeq 1$
mag.

\begin{figure}[t]
\includegraphics[angle=270,width=\columnwidth]{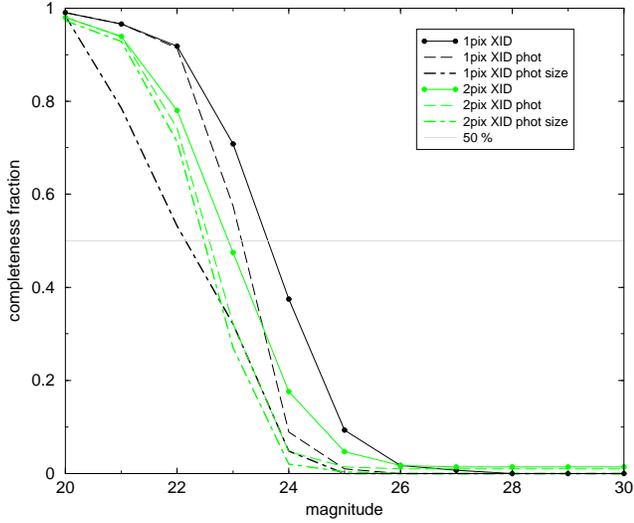}
\caption{Completeness fractions for artificial clusters with FWHM = $\simeq$ 10 pc
(= 1pixel) and FWHM = $\simeq$ 20 pc (= 2pixel) for a variety of selection criteria (see text for
details)} 
\label{fig:complete2}
\end{figure}

Computing completeness functions for 2 regions (characterised by
different background and cluster density levels) and 2 sizes
(characteristic for our observed cluster sample) and imposing the size
restriction onto the artificial clusters (labelled "XID photo size") 
allowed us to attribute realistic completeness fractions to each
individual cluster, strengthening the subsequent analysis of the LFs.
Imposing the size restrictions decrease the completeness by up to
$\simeq 1$ mag, while different cluster sizes and regions within the
galaxies contribute completeness changes of $\simeq 0.6$ and $\simeq
1$ mag, respectively. This very complex dependence of the completeness
fractions on a number of input parameters clearly shows the necessity
to {\bf model the completeness functions as realistically as possible,
taking into account all cluster selection criteria.}  For the same
region as in Fig. \ref{fig:complete}, results for different cluster
sizes and selection criteria are shown in Fig. \ref{fig:complete2}.

For additional details of the completeness determination, see
\cite{Anders07}.

\section{The model to analyse the luminosity functions}
\label{sec:model}

In collaboration with statisticians from the University of G\"ottingen we developed
a suite of statistical tools. This allows to fit the LFs of our
young star clusters with either a Gaussian distribution or a power-law, access the
accuracy of the fits and the fitted parameters, and establish the statistical
significance of differences in the goodness-of-fit between the two fitted test
distributions.

The fitting of a observed LF takes the photometric error and
appropriate completeness fraction of each individual cluster into account, and
independently fits either a Gaussian distribution or a power-law to the LF, yielding a set of parameters characterising the best-fitting model and a
likelihood parameter for either test distribution.

We find a strong superiority (measured by the ratio of the likelihoods
for Gaussian and power-law fits) of the Gaussian fit compared to the
power-law fit (see Sect. \ref{sec:results}). To validate this result,
we have to test its statistical significance, i.e. to determine the
probability to achieve such strong superiority of the Gaussian
distribution if the {\sl true underlying} LF is a power-law, distorted
by the error distribution and the completeness function. In a
Monte-Carlo approach we draw artificial observations from the
best-fitting power-law, use our statistics tools to fit them again
with Gaussian and power-law distribution, and determine the likelihood
ratio for these artificial observations. The likelihood ratio of the
observed distribution compared to the distribution of likelihood
ratios from our artificial tests then yields the probability that the
determined superiority of the Gaussian fits is consistent with an
underlying power-law distribution.

The uncertainties of the best-fitting model parameters are estimated by
bootstrapping.

For additional details of the statistical model, see \cite{Anders07}.

\section{Results}
\label{sec:results}

\begin{figure}[t]
\begin{center}
\includegraphics[angle=0,width=0.75\columnwidth]{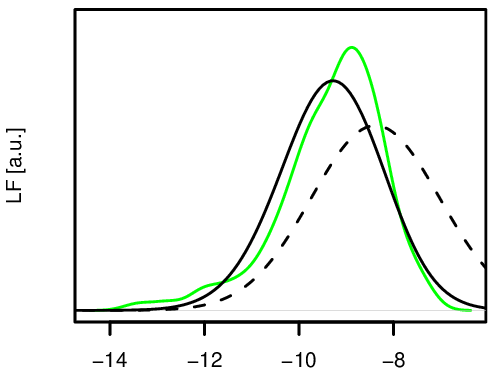}
\includegraphics[angle=0,width=0.75\columnwidth]{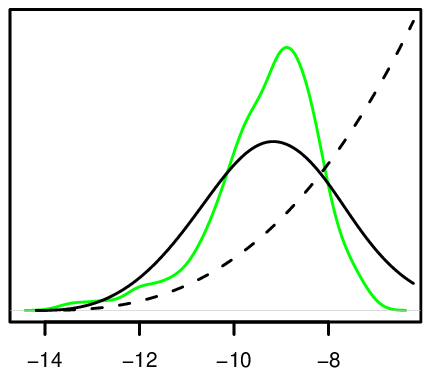}
\caption{V-band luminosity functions for our Antennae clusters and fit: Observed
(green solid line), best fit (black solid) and underlying distribution of best fit
(black dashed). Fitted test distributions are: Gaussian (left panel) and power-law
(right panel)} 
\label{fig:fits}
\end{center}
\end{figure}

As example, the LF in the V band for our YSC sample in the Antennae
galaxies is shown in Fig. \ref{fig:fits}. From visual inspection, the best fit with
an underlying Gaussian distribution appears to represent the data better than the
best fit with an underlying power-law distribution.

To quantify this, we perform our likelihood ratio test and the Monte-Carlo analysis.
For the observed distributions we find a likelihood ratio value of 23.1 (where
larger values are equivalent to a stronger superiority of the Gaussian fit when
compared to the power-law fit). The Monte-Carlo analysis with 1000 test
realisations drawn from the best-fitting power-law distribution results in a maximum
likelihood ratio of 11.4, hence {\bf none} of the test distributions can reproduce a
superiority of the Gaussian distribution as strong as observed. Utilising the
properties of Bernoulli-distributed variables, this corresponds to a probability $<$
0.5\% that the underlying distribution is still consistent with a power-law.

This result is valid also for the other bands (except the I band, for
which the observations are significantly shallower), and for several
age and size subsamples. For more details, see \cite{Anders07}.

\acknowledgments
We would like to thank Nicolai Bissantz and Leif Boysen for their constant help and
development of the statistics tools. 



\bibliographystyle{Spr-mp-nameyear-cnd}  
\bibliography{Anders}  				  

\begin{thebibliography}{}
\ifx \bisbn   \undefined \def \bisbn  #1{ISBN #1}   \fi
\ifx \binits  \undefined \def \binits#1{#1} \fi
\ifx \bauthor  \undefined \def \bauthor#1{#1} \fi
\ifx \batitle  \undefined \def \batitle#1{#1} \fi
\ifx \bjtitle  \undefined \def \bjtitle#1{#1} \fi
\ifx \bvolume  \undefined \def \bvolume#1{\textbf {#1}} \fi
\ifx \byear  \undefined \def \byear#1{#1} \fi
\ifx \bissue  \undefined \def \bissue#1{#1} \fi
\ifx \bfpage  \undefined \def \bfpage#1{#1} \fi
\ifx \blpage  \undefined \def \blpage #1{#1} \fi
\ifx \burl  \undefined \def \burl#1{#1} \fi
\ifx \doiurl  \undefined \def \doiurl#1{#1} \fi
\ifx \betal  \undefined \def \betal#1{#1} \fi
\ifx \binstitutionaled  \undefined \def \binstitutionaled#1{#1} \fi
\ifx \binstitute  \undefined \def \binstitute#1{#1} \fi
\ifx \bctitle  \undefined \def \bctitle#1{#1} \fi
\ifx \beditor  \undefined \def \beditor#1{#1} \fi
\ifx \bpublisher  \undefined \def \bpublisher#1{#1} \fi
\ifx \bbtitle  \undefined \def \bbtitle#1{#1} \fi
\ifx \bedition  \undefined \def \bedition#1{#1} \fi
\ifx \bseriesno  \undefined \def \bseriesno#1{#1} \fi
\ifx \blocation  \undefined \def \blocation#1{#1} \fi
\ifx \bsertitle  \undefined \def \bsertitle#1{#1} \fi
\ifx \bsnm \undefined \def \bsnm#1{#1} \fi
\ifx \bsuffix \undefined \def \bsuffix#1{#1} \fi
\ifx \bparticle \undefined \def \bparticle#1{#1} \fi
\ifx \barticle \undefined \def \barticle#1{#1} \fi
\ifx \botherref \undefined \def \botherref #1{#1} \fi
\ifx \url \undefined \def \url#1{#1} \fi
\ifx \bchapter \undefined \def \bchapter#1{#1} \fi
\ifx \bbook \undefined \def \bbook#1{#1} \fi
\ifx \bcomment \undefined \def \bcomment#1{#1} \fi
\ifx \oauthor \undefined \def \oauthor#1{#1} \fi
\def \endbibitem {}

\bibitem[\protect\citeauthoryear{{Anders} et~al.}{2007}]{Anders07}
\begin{barticle}
\bauthor{\bsnm{{Anders}},~\binits{P.}},
  \bauthor{\bsnm{{Bissantz}},~\binits{N.}},
  \bauthor{\bsnm{{Boysen}},~\binits{L.}}, \bauthor{\bsnm{{de
  Grijs}},~\binits{R.}}, \bauthor{\bsnm{{Fritze-v.~Alvensleben}},~\binits{U.}}:
\bjtitle{\mnras} \bvolume{377}, \bfpage{91} (\byear{2007})
\end{barticle}
\endbibitem

\bibitem[\protect\citeauthoryear{{Anders} and
  {Fritze-v.~Alvensleben}}{2003}]{2003A&A...401.1063A}
\begin{barticle}
\bauthor{\bsnm{{Anders}},~\binits{P.}},
  \bauthor{\bsnm{{Fritze-v.~Alvensleben}},~\binits{U.}}:
\bjtitle{\aap} \bvolume{401}, \bfpage{1063} (\byear{2003})
\end{barticle}
\endbibitem

\bibitem[\protect\citeauthoryear{{Anders} et~al.}{2006}]{2006A&A...451..375A}
\begin{barticle}
\bauthor{\bsnm{{Anders}},~\binits{P.}}, \bauthor{\bsnm{{Gieles}},~\binits{M.}},
  \bauthor{\bsnm{{de Grijs}},~\binits{R.}}:
\bjtitle{\aap} \bvolume{451}, \bfpage{375} (\byear{2006})
\end{barticle}
\endbibitem

\bibitem[\protect\citeauthoryear{{Ashman} and
  {Zepf}}{1998}]{1998gcs..book.....A}
\begin{bbook}
\bauthor{\bsnm{{Ashman}},~\binits{K.M.}},
  \bauthor{\bsnm{{Zepf}},~\binits{S.E.}}:
\bbtitle{{Globular Cluster Systems}}. \bsertitle{{Cambridge astrophysics series
  ; 30}}. \bpublisher{{Cambridge University Press}}, \blocation{{Cambridge,
  U.~K.~; New York}} (\byear{1998})
\end{bbook}
\endbibitem

\bibitem[\protect\citeauthoryear{{Bruzual} and
  {Charlot}}{2003}]{2003MNRAS.344.1000B}
\begin{barticle}
\bauthor{\bsnm{{Bruzual}},~\binits{G.}},
  \bauthor{\bsnm{{Charlot}},~\binits{S.}}:
\bjtitle{\mnras} \bvolume{344}, \bfpage{1000} (\byear{2003})
\end{barticle}
\endbibitem

\bibitem[\protect\citeauthoryear{{de Grijs} and
  {Anders}}{2006}]{2006MNRAS.366..295D}
\begin{barticle}
\bauthor{\bsnm{{de Grijs}},~\binits{R.}},
  \bauthor{\bsnm{{Anders}},~\binits{P.}}:
\bjtitle{\mnras} \bvolume{366}, \bfpage{295} (\byear{2006})
\end{barticle}
\endbibitem

\bibitem[\protect\citeauthoryear{{de Grijs}
  et~al.}{2003b}]{2003MNRAS.343.1285D}
\begin{barticle}
\bauthor{\bsnm{{de Grijs}},~\binits{R.}},
  \bauthor{\bsnm{{Anders}},~\binits{P.}},
  \bauthor{\bsnm{{Bastian}},~\binits{N.}},
  \bauthor{\bsnm{{Lynds}},~\binits{R.}},
  \bauthor{\bsnm{{Lamers}},~\binits{H.J.G.L.M.}},
  \bauthor{\bsnm{{O'Neil}},~\binits{E.J.}}:
\bjtitle{\mnras} \bvolume{343}, \bfpage{1285} (\byear{2003b})
\end{barticle}
\endbibitem

\bibitem[\protect\citeauthoryear{{de Grijs}
  et~al.}{2003a}]{2003NewA....8..155D}
\begin{barticle}
\bauthor{\bsnm{{de Grijs}},~\binits{R.}},
  \bauthor{\bsnm{{Lee}},~\binits{J.T.}}, \bauthor{\bsnm{{Clemencia Mora
  Herrera}},~\binits{M.}},
  \bauthor{\bsnm{{Fritze-v.~Alvensleben}},~\binits{U.}},
  \bauthor{\bsnm{{Anders}},~\binits{P.}}:
\bjtitle{New Astronomy} \bvolume{8}, \bfpage{155} (\byear{2003a})
\end{barticle}
\endbibitem

\bibitem[\protect\citeauthoryear{{Fritze-v.~Alvensleben}}{1999}]{1999A&A...342%
L..25F}
\begin{barticle}
\bauthor{\bsnm{{Fritze-v.~Alvensleben}},~\binits{U.}}:
\bjtitle{\aap} \bvolume{342}, \bfpage{L25} (\byear{1999})
\end{barticle}
\endbibitem

\bibitem[\protect\citeauthoryear{{Goudfrooij}
  et~al.}{2004}]{2004ApJ...613L.121G}
\begin{barticle}
\bauthor{\bsnm{{Goudfrooij}},~\binits{P.}},
  \bauthor{\bsnm{{Gilmore}},~\binits{D.}},
  \bauthor{\bsnm{{Whitmore}},~\binits{B.C.}},
  \bauthor{\bsnm{{Schweizer}},~\binits{F.}}:
\bjtitle{\apjl} \bvolume{613}, \bfpage{L121} (\byear{2004})
\end{barticle}
\endbibitem

\bibitem[\protect\citeauthoryear{{Harris}}{1991}]{1991ARA&A..29..543H}
\begin{barticle}
\bauthor{\bsnm{{Harris}},~\binits{W.E.}}:
\bjtitle{\araa} \bvolume{29}, \bfpage{543} (\byear{1991})
\end{barticle}
\endbibitem

\bibitem[\protect\citeauthoryear{{Hunter} et~al.}{2003}]{2003AJ....126.1836H}
\begin{barticle}
\bauthor{\bsnm{{Hunter}},~\binits{D.A.}},
  \bauthor{\bsnm{{Elmegreen}},~\binits{B.G.}},
  \bauthor{\bsnm{{Dupuy}},~\binits{T.J.}},
  \bauthor{\bsnm{{Mortonson}},~\binits{M.}}:
\bjtitle{\aj} \bvolume{126}, \bfpage{1836} (\byear{2003})
\end{barticle}
\endbibitem

\bibitem[\protect\citeauthoryear{{Larsen}}{1999}]{1999A&AS..139..393L}
\begin{barticle}
\bauthor{\bsnm{{Larsen}},~\binits{S.S.}}:
\bjtitle{\aaps} \bvolume{139}, \bfpage{393} (\byear{1999})
\end{barticle}
\endbibitem

\bibitem[\protect\citeauthoryear{{Leitherer}
  et~al.}{1999}]{1999ApJS..123....3L}
\begin{barticle}
\bauthor{\bsnm{{Leitherer}},~\binits{C.}},
  \bauthor{\bsnm{{Schaerer}},~\binits{D.}},
  \bauthor{\bsnm{{Goldader}},~\binits{J.D.}},
  \bauthor{\bsnm{{Delgado}},~\binits{R.M.G.}},
  \bauthor{\bsnm{{Robert}},~\binits{C.}},
  \bauthor{\bsnm{{Kune}},~\binits{D.F.}}, \bauthor{\bsnm{{de
  Mello}},~\binits{D.F.}}, \bauthor{\bsnm{{Devost}},~\binits{D.}},
  \bauthor{\bsnm{{Heckman}},~\binits{T.M.}}:
\bjtitle{\apjs} \bvolume{123}, \bfpage{3} (\byear{1999})
\end{barticle}
\endbibitem

\bibitem[\protect\citeauthoryear{{Meurer}}{1995}]{1995Natur.375..742M}
\begin{barticle}
\bauthor{\bsnm{{Meurer}},~\binits{G.R.}}:
\bjtitle{\nat} \bvolume{375}, \bfpage{742} (\byear{1995})
\end{barticle}
\endbibitem

\bibitem[\protect\citeauthoryear{{Schweizer} and
  {Seitzer}}{1998}]{1998AJ....116.2206S}
\begin{barticle}
\bauthor{\bsnm{{Schweizer}},~\binits{F.}},
  \bauthor{\bsnm{{Seitzer}},~\binits{P.}}:
\bjtitle{\aj} \bvolume{116}, \bfpage{2206} (\byear{1998})
\end{barticle}
\endbibitem

\bibitem[\protect\citeauthoryear{{van den Bergh} and
  {Lafontaine}}{1984}]{1984AJ.....89.1822V}
\begin{barticle}
\bauthor{\bsnm{{van den Bergh}},~\binits{S.}},
  \bauthor{\bsnm{{Lafontaine}},~\binits{A.}}:
\bjtitle{\aj} \bvolume{89}, \bfpage{1822} (\byear{1984})
\end{barticle}
\endbibitem

\bibitem[\protect\citeauthoryear{{Whitmore} et~al.}{1999}]{1999AJ....118.1551W}
\begin{barticle}
\bauthor{\bsnm{{Whitmore}},~\binits{B.C.}},
  \bauthor{\bsnm{{Zhang}},~\binits{Q.}},
  \bauthor{\bsnm{{Leitherer}},~\binits{C.}},
  \bauthor{\bsnm{{Fall}},~\binits{S.M.}},
  \bauthor{\bsnm{{Schweizer}},~\binits{F.}},
  \bauthor{\bsnm{{Miller}},~\binits{B.W.}}:
\bjtitle{\aj} \bvolume{118}, \bfpage{1551} (\byear{1999})
\end{barticle}
\endbibitem

\end{thebibliography}

%

\end{document}